\documentclass[aps,twocolumn,prl,floatfix,amsmath,amssymb,superscriptaddress]{revtex4}

\usepackage{epsfig}
\usepackage{graphicx}
\usepackage{dcolumn}
\usepackage{bm}
\usepackage{color}

\begin{document}

\title{Scanning tunneling spectroscopy of layers of superconducting 2H-TaSe$_\textbf{2}$: Evidence for a zero bias anomaly in single layers}

\author{J.A. Galvis}
\affiliation{Laboratorio de Bajas Temperaturas, Departamento de F\'isica de la Materia Condensada \\ Instituto de Ciencia de Materiales Nicol\'as Cabrera, Facultad de Ciencias \\ Universidad Aut\'onoma de Madrid, E-28049 Madrid, Spain}
\author{P.\, Rodi\`ere}
\affiliation{Institut N\'eel, CRS/UJF, 25, Av. des Martyrs, BP166, 38042 Grenoble Cedex 9, France}
\author{I. Guillamon}
\affiliation{Laboratorio de Bajas Temperaturas, Departamento de F\'isica de la Materia Condensada \\ Instituto de Ciencia de Materiales Nicol\'as Cabrera, Facultad de Ciencias \\ Universidad Aut\'onoma de Madrid, E-28049 Madrid, Spain}
\affiliation{H.H. Wills Physics Laboratory, University of Bristol, Tyndall Avenue, Bristol, BS8 1TL, UK}
\author{M.R. Osorio}
\affiliation{Laboratorio de Bajas Temperaturas, Departamento de F\'isica de la Materia Condensada \\ Instituto de Ciencia de Materiales Nicol\'as Cabrera, Facultad de Ciencias \\ Universidad Aut\'onoma de Madrid, E-28049 Madrid, Spain}
\author{J.G. Rodrigo}
\affiliation{Laboratorio de Bajas Temperaturas, Departamento de F\'isica de la Materia Condensada \\ Instituto de Ciencia de Materiales Nicol\'as Cabrera, Facultad de Ciencias \\ Universidad Aut\'onoma de Madrid, E-28049 Madrid, Spain}
\author{L.\, Cario}
\affiliation{Institut des Mat\'eriaux Jean Rouxel (IMN), Universit\'e de Nantes, CNRS, 2 rue de la Houssini\'ere, BP 32229, 44322 Nantes Cedex 03, France}
\author{E. Navarro-Moratalla}
\affiliation{Instituto de Ciencia Molecular (ICMol), Universidad de Valencia, Catedr\'atico Jos\'e Beltr\'an 2, 46980 Paterna, Spain}
\author{E. Coronado}
\affiliation{Instituto de Ciencia Molecular (ICMol), Universidad de Valencia, Catedr\'atico Jos\'e Beltr\'an 2, 46980 Paterna, Spain}
\author{S. Vieira}
\affiliation{Laboratorio de Bajas Temperaturas, Departamento de F\'isica de la Materia Condensada \\ Instituto de Ciencia de Materiales Nicol\'as Cabrera, Facultad de Ciencias \\ Universidad Aut\'onoma de Madrid, E-28049 Madrid, Spain}
\author{H. Suderow$^*$}
\affiliation{Laboratorio de Bajas Temperaturas, Departamento de F\'isica de la Materia Condensada \\ Instituto de Ciencia de Materiales Nicol\'as Cabrera, Facultad de Ciencias \\ Universidad Aut\'onoma de Madrid, E-28049 Madrid, Spain}

\begin{abstract}
We report a characterization of surfaces of the dichalcogenide TaSe$_2$ using scanning tunneling microscopy and spectroscopy (STM/S) at 150 mK. When the top layer has the 2H structure and the layer immediately below the 1T structure, we find a singular spatial dependence of the tunneling conductance below 1 K, changing from a zero bias peak on top of Se atoms to a gap in between Se atoms. The zero bias peak is additionally modulated by the commensurate  $3a_0 \times 3a_0$ charge density wave of 2H-TaSe$_2$. Multilayers of 2H-TaSe$_2$ show a spatially homogeneous superconducting gap with a critical temperature also of 1 K. We discuss possible origins for the peculiar tunneling conductance in single layers.
\end{abstract}

\maketitle

\section{Introduction.}

\begin{figure}
\includegraphics[width=0.48\textwidth,clip]{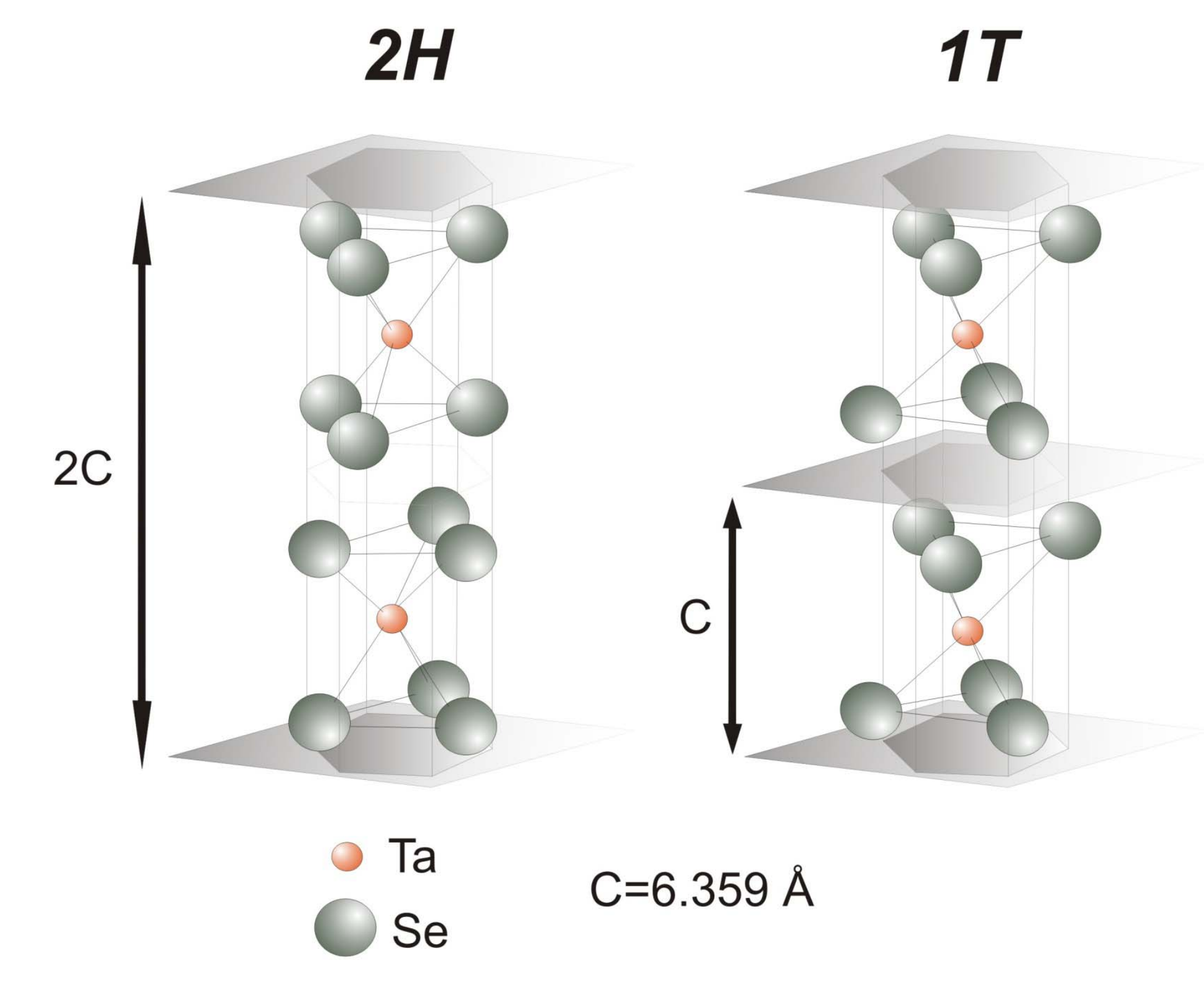}
\caption{We show two possible polytypes of TaSe$_2$, the 2H and 1T phases, which consist of two TaSe$_2$ hexagonal prisms rotated with respect to each other, and one trigonal prism. Se atoms are shown as grey spheres and Ta atoms as smaller red spheres. The coordination of Ta is trigonal prismatic in the 2H structure, and octahedral in the 1T structure.}
\label{structure}
\end{figure}

There is a rather general interest arising in transition metal dichalchogenides because superconductivity, charge density wave (CDW), quantum criticality and single layer physics are found in just a few compounds\cite{Sacks98,CastroNeto01,NatureNanotech12,Vicent1980,Feng2012,Ye12,Chubukov12}. These systems share the formula MX$_2$, where M is a transition metal such as for instance Nb or Ta, and X a chalchogen such as Se or S. The crystallographic structure consists of layers with small coupling among them. In TaSe$_2$, layers are made of Se-Ta-Se units with two triangular sheets of Se atoms separated by one sheet of Ta atoms. Interlayer bonding is weak through van der Waals forces. The relative arrangement of the Se triangles varies in different polytypes of the same compound. In Fig.\ref{structure} we show the structure of the the 2H and the 1T polytypes. 2H polytype has a unit cell consisting of two TaSe$_2$ units, each built up by two Se triangles separated by Ta atoms. N
 ote the 30$^\circ$ rotation between the Se triangles of the two prisms in 2H-TaSe$_2$. The unit cell of the 1T polytype is composed of a single layer of TaSe$_2$, with the Se triangles rotated to each other by 30$^\circ$\cite{Wilson75}.

\vskip 0.5cm

\begin{figure*}
\includegraphics[width=0.9\textwidth,clip]{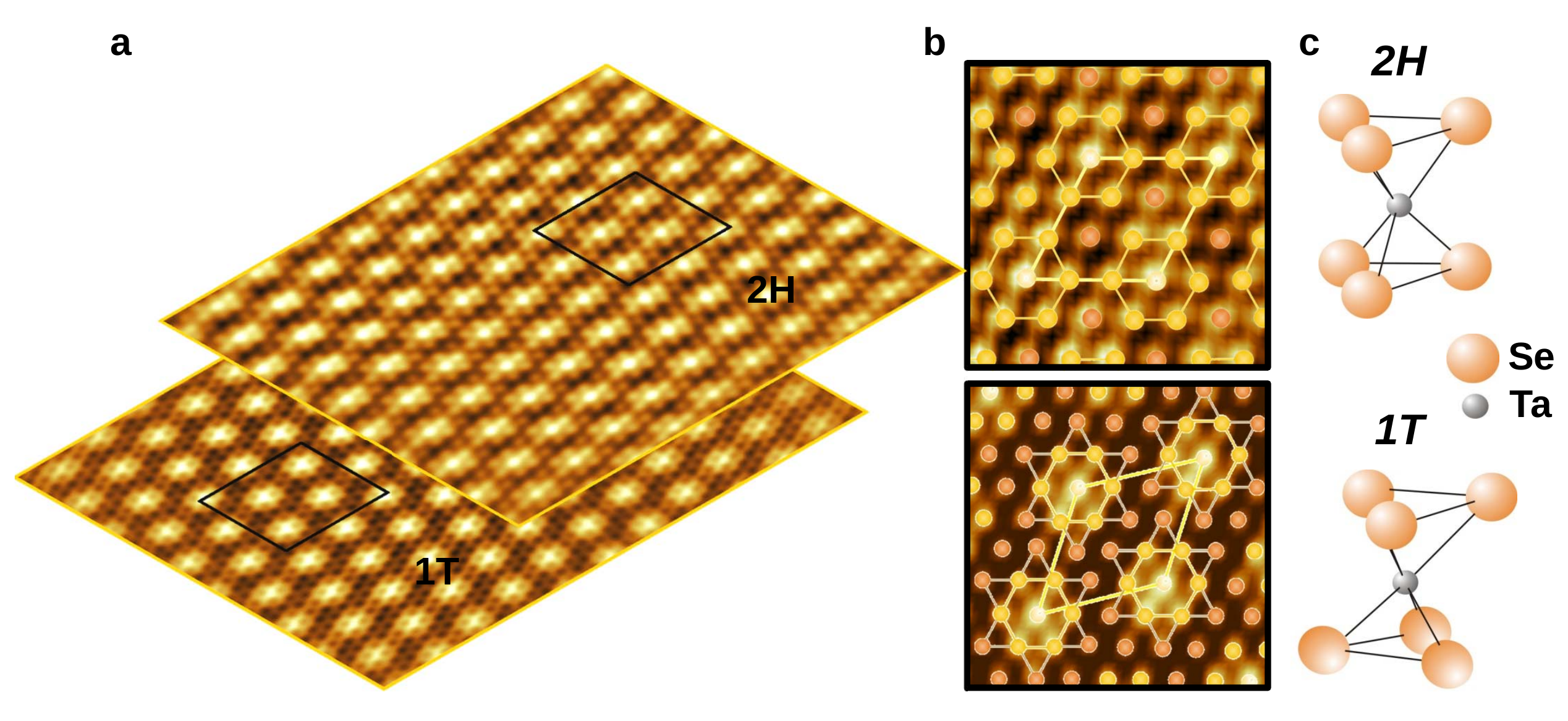}
\caption{We show a schematical representation of scanning tunneling microscopy (STM) images of the atomic arrangements discussed. We highlight the top sheet as 2H. It shows a $3a_0 \times 3a_0$ superlattice modulated charge density wave (CDW). The layer immediately below, highlighted as 1T has a different Moir\'e pattern with a $\sqrt{13}a_0 \times \sqrt{13}a_0$ CDW. In 2H-TaSe$_2$, top and bottom Se atoms are aligned (top figure in c). In 1T-TaSe$_2$, the two Se triangles are rotated (bottom figure in c). The middle panels (b) show in more detail the CDW patterns found in each area. There are three inequivalent atomic sites in the pattern, as highlighted by yellow, orange and white points. The incommensurate CDW $\sqrt{13}a_0 \times \sqrt{13}a_0$ pattern shown in the bottom panel of (b) consists of stars of David, with also three inequivalent sites. The arrangement shown in (a), with one 2H-TaSe$_2$ sheet on top of 1T-TaSe$_2$ presents the singular superconducting features d
 iscussed in the text.}
\label{picture}
\end{figure*}

The c-axis interlayer coupling decreases from 2H-NbSe$_2$, 2H-TaS$_2$ to 2H-TaSe$_2$. Superconductivity in 2H-NbSe$_2$ (T$_c$=7 K) is multiband and s-wave, with a strong in-plane anisotropy related to the charge density wave\cite{Sacks98,CastroNeto01,Rodrigo04PhysC,Guillamon08}. 2H-TaS$_2$  (T$_c$=0.8 K) shows anisotropic charge density wave patterns at the surface. These anisotropic patterns were first proposed indicative of a chiral charge density wave\cite{Ishioka10, Guillamon11}. Subsequent analysis showed instead that these patterns are due to polar charge and orbital order\cite{Wezel12}. The relationship between superconductivity and such in-plane anisotropic charge order is yet unclear.

There are few reports about superconductivity in 2H-TaSe$_2$, showing superconducting diamagnetic signals and zero resistance at temperatures below 150 mK \cite{Wilson75,Kumakura96,Yokota00}. Zero temperature extrapolated critical fields are very low (1.4 mT with the field along the c-axis). On the other hand, calculations and angular resolved photoemission have unveiled the Fermi surface and band structure in detail, and in particular their relationship with the onset of CDW\cite{Wilson75,Rossnagel11}. Charge order in 2H-TaSe$_2$ is incommensurate between 122 K and 90 K, below which it locks into a $3a_0\times3a_0$ (with $a_0$ being the in plane lattice parameter) commensurate charge modulation\cite{Slough86,Rossnagel11}. The incommensurate CDW has been shown to lead to incomplete gapping. Its form and temperature dependence has been discussed in relationship to the opening of the pseudogap in cuprates\cite{Rossnagel11}. 1T-TaSe$_2$ also develops a charge density wave at low
  temperatures\cite{Wilson75,Bulaevskii76}, with, however, a periodicity of $\sqrt{13}a_0 \times \sqrt{13}a_0$ rotated by $13.5^\circ$ with respect to the atomic lattice\cite{Slough86,Giambattista88}. Within the charge ordered state, 1T-TaSe$_2$ has a high in-plane resistivity and a c-axis mean free path below the interplanar distance. No superconductivity has been reported in this material so far at ambient pressure and without doping. One of the most direct methods to determine the polytype of the surface is to measure the charge modulations using Scanning Tunneling Microscopy (STM, see Fig.\ref{picture}). Here we study 2H-TaSe$_2$ using STM at 150 mK and demonstrate that repeated exfoliation gives surfaces with single layer crystals of 2H-TaSe$_2$ on top of 1T-TaSe$_2$ (Fig.\ref{picture}). Superconductivity with a strongly enhanced critical temperature with respect to the bulk is also observed, and single layers show highly anomalous tunneling conductance features.

\section{Experimental}

\begin{figure}
\includegraphics[width=0.46\textwidth,clip]{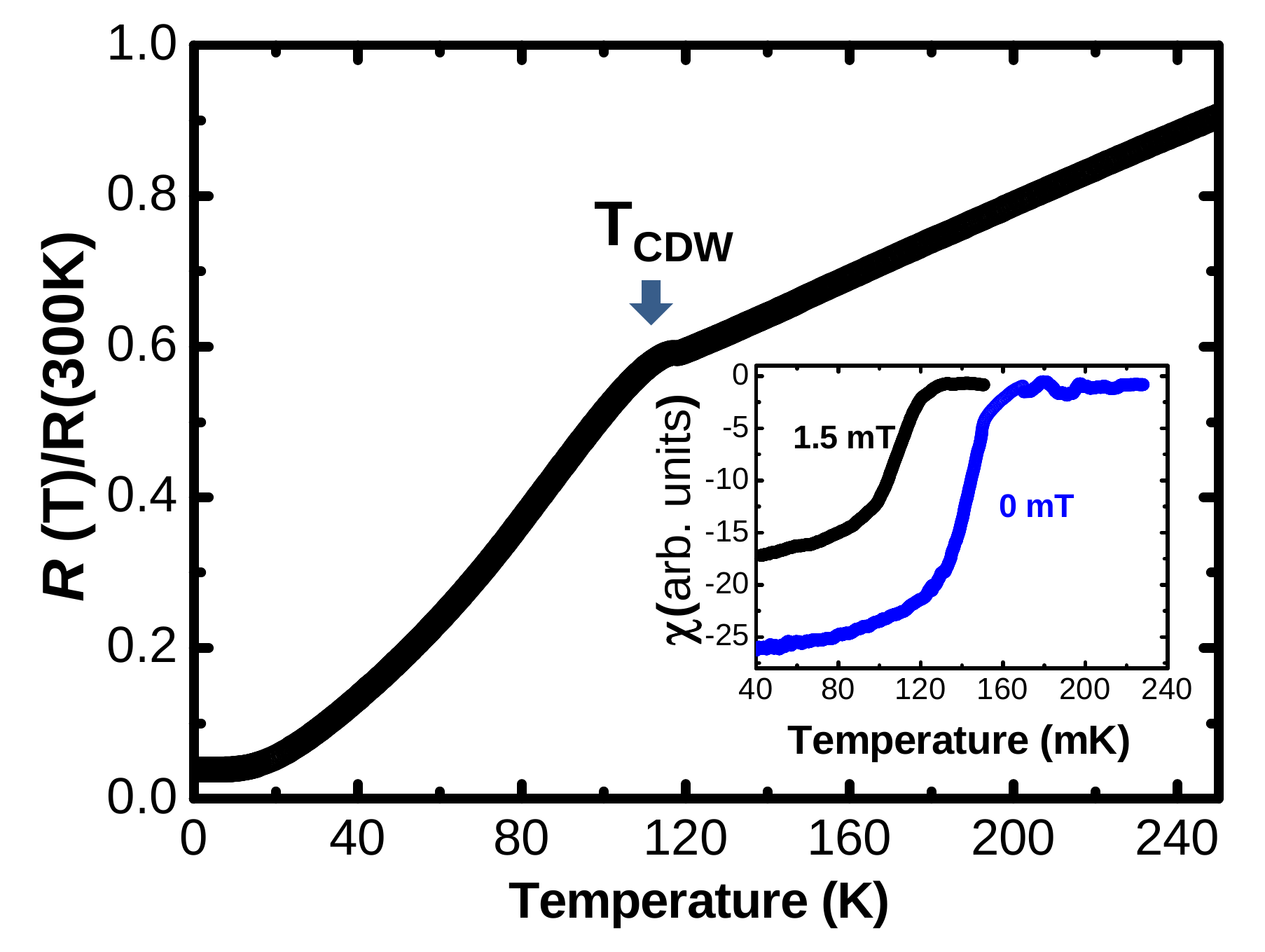}
\caption{The resistance vs temperature of our samples normalized to its ambient temperature value is shown in the main panel down to 0.5 K. Residual resistivity ratio is of 26. Onset of charge density wave (CDW) order is shown by an arrow. The temperature dependent susceptibility, which shows the superconducting transition, is shown in the inset.}
\label{transport}
\end{figure}

\begin{figure}
\includegraphics[width=0.46\textwidth,clip]{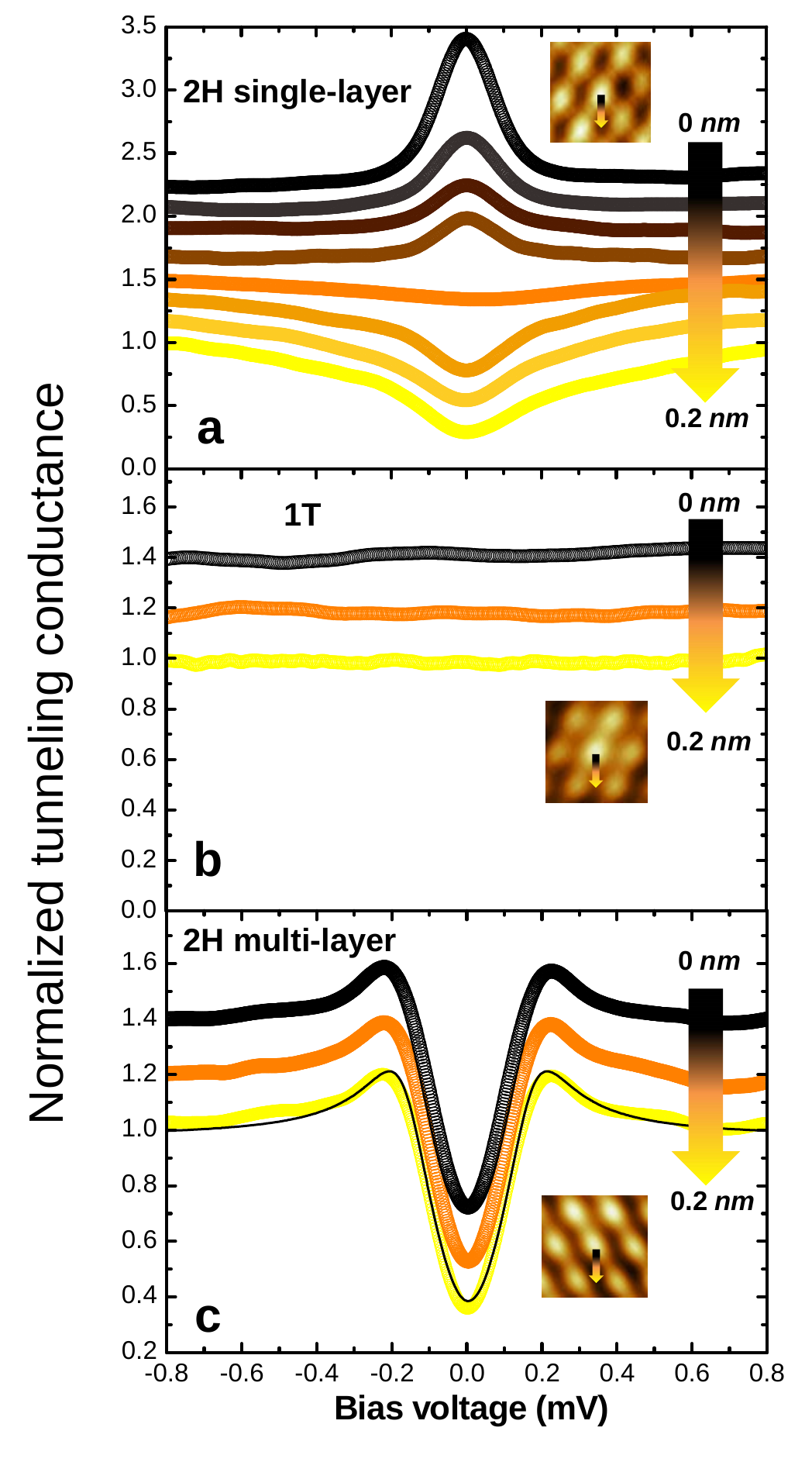}
\caption{Tunneling conductance versus bias voltage in single layer 2H-TaSe$_2$ on top of 1T-TaSe$_2$ (a), as in Fig.\ref{picture}, in 1T-TaSe$_2$  (b) and in multilayer 2H-TaSe$_2$ (c) taken at 0.15 K. The black line on top of the yellow curve in c is a fit to superconducting gap as described in the discussion section. Different tunneling conductance curves are taken when the tip moves from a Se atom (black) to an intersite (yellow). Insets show the path on atomic size topography images involving a hexagon of Se atoms. Following other paths of the six-fold symmetry lead to qualitatively the same features.}
\label{tunnel}
\end{figure}

\begin{figure*}
\includegraphics[width=0.9\textwidth,clip]{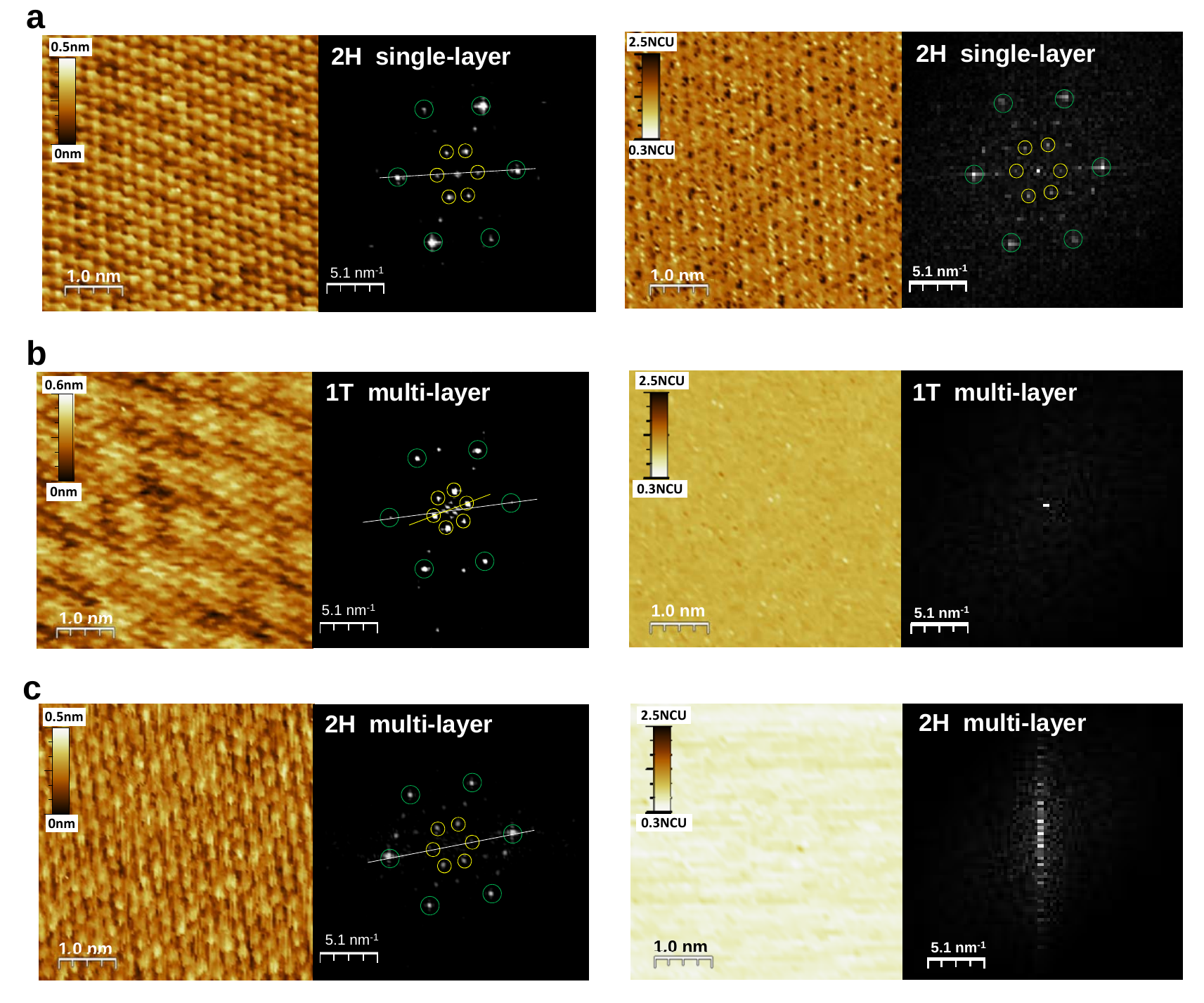}
\caption{Scanning tunneling microscopy (left panels) and zero bias conductance maps (right panels) images in single layer 2H-TaSe$_2$ on top of 1T-TaSe$_2$ (a), as in Fig.\ref{picture}, in 1T-TaSe$_2$  (b) and in multilayer 2H-TaSe$_2$ (c) taken at 0.15 K.  Real space (left) and Fourier transforms (right) are shown together in each panel. Green circles show the position of the Bragg peaks due to the atomic Se modulation and yellow circles show the peaks due to the CDW modulation. Lines are shown to highlight the angular difference between atomic and CDW modulations. The color scale of the zero bias conductance maps is given in normalized conductance units (NCU), that is, the zero bias conductance normalized to the conductance above 1 mV. Color scale of the upper panels corresponds to the colors used in the upper panel of Fig.\ref{tunnel}. Black points correspond to curves similar to the black tunneling conductance curves shown in Fig.\ref{tunnel}, and yellow points to yellow 
 tunneling conductance curves. Images are unfiltered.}
\label{summary}
\end{figure*}

We measure TaSe$_2$ in a home built low noise STM arrangement installed in a dilution refrigerator, which is capable of cooling down the microscope to about 100 mK. We can make tunneling conductance curves with a resolution of 20 $\mu$V, which corresponds to an energy resolution above 150 mK. The STM has a sample holder that allows us to change \emph{in-situ} the scanning window\cite{Suderow11}. We use a tip of Au, which we clean through repeated indentation on an Au cleaning pad\cite{Rodrigo04b,Suderow11}. The samples are 2H-TaSe$_2$ crystals grown using iodine vapour-transport from stoichiometric prereacted powders. Bulk susceptibility measurements of our samples are shown in the inset of Fig.\ref{transport} for two magnetic fields. Susceptibility gives a diamagnetic signal with the same T$_c$ as in previous work (150 mK)\cite{Wilson75,Kumakura96,Yokota00}. The temperature dependence of the resistance normalized at ambient temperature is shown in the main panel of Fig.\ref{
 transport} and was measured between 300 K and 0.5 K. CDW onset causes a change in the resistance near T$_{CDW}$=122 K. The residual resistivity ratio of the TaSe$_2$ samples is of 26, implying samples similar to those used to observe quantum oscillations and measure the Fermi surface\cite{Hillenius78,Fleming77,Graebner76}. We glued the 2H-TaSe$_2$ sample using silver epoxy onto the sample holder. We used scotch to remove the upper layers of 2H-TaSe$_2$ at ambient conditions, and optically inspected the result prior to inserting the set-up into liquid helium. The scotch cleaving procedure works nicely, but it is more involved than cleaving 2H-NbSe$_2$\cite{Rodrigo04b,Guillamon08,Rodrigo04PhysC}. In 2H-NbSe$_2$, it is easy to obtain a clean looking shiny surface without free standing sheets. But in our samples of 2H-TaSe$_2$, the cleaved surface nearly always consists of big loose sheets which have to be manually removed using tweezers, until we observe a flat and shiny surfac
 e. We cooled down samples eleven times, making each time a new cleave. In five cool downs, we observed pure 1T-TaSe$_2$ surface with the $\sqrt{13}\times\sqrt{13}$ CDW, without any traces of superconductivity. In two cool downs, we found 2H-TaSe$_2$ behavior over the whole surface without any trace of superconductivity. This is what we expect for bulk 2H-TaSe$_2$ with the above mentioned energy resolution, above T$_c$ of 150 mK. In four cool downs, we observed the phenomena described below with mixed hexagonal (2H) and trigonal (1T) surfaces. In each of them, we studied about ten different 2$\times$2 $\mu$m scanning windows, finding the behavior discussed below. 

Typically, we scan with a set-point at bias voltages of about 2 mV and a tunneling conductance of 1$\mu S$ or below. We did not find a significant dependence of the tunneling curves or the images on the set-point tunneling conductance within an order of magnitude above or below this value. I-V curves are numerically derivated, as in previous work\cite{Guillamon08,Rodrigo04b}. We normalize to the conductance value at 1 mV, and curves are flat between 1 mV and 2 mV. We did not systematically study curves above this bias voltage value. Images have been made using home made data acquisition software, and rendered using home-made, WSxM\cite{Horcas07} and matlab programs.

\section{Results}

The STM topography shows the atomic Se lattice and either the $3a_0\times3a_0$ CDW with the same orientation to the Se lattice and commensurate to it, or Moir\'e patterns characteristic of a CDW with a periodicity of $\sqrt{13}a_0 \times \sqrt{13}a_0$ rotated by $13.5^\circ$ with respect to the atomic Se lattice. The first identifies 2H-TaSe$_2$ surfaces, and the second 1T-TaSe$_2$ surfaces (Fig.\ref{picture}). We find atomically flat 2H-TaSe$_2$ layers, showing immediately below a surface of 1T-TaSe$_2$. Steps between both surfaces are of 1.2 nm, corresponding to two layers of the sandwich Se-Ta-Se. Thus, the upper layers are single layer crystals of 2H-TaSe$_2$. We also find 1T-TaSe$_2$ or 2H-TaSe$_2$ surfaces over large areas and many different steps. These correspond to multilayer 1T-TaSe$_2$ and 2H-TaSe$_2$, respectively. We have made a thorough spectroscopy and microscopy characterization of all observed surfaces by taking simultaneously topography and tunneling conduct
 ance (Figs.\ref{tunnel} and \ref{summary}). In Fig.\ref{tunnel} we show curves obtained when scanning from the top of a Se atom to an intersite at the center of three Se atoms. On surfaces of 2H-TaSe$_2$ on top of 1T-TaSe$_2$, a clear zero bias peak is observed on top of the Se atoms (Figs.\ref{tunnel}a and \ref{summary}a). The zero bias peak evolves continously into a V-shape conductance at the intersites, following the topography. This is made clear in the Fourier transform of the zero bias conductance map which shows Bragg peaks at the positions corresponding to atomic and CDW reciprocal vectors. 

In the 1T-TaSe$_2$ layers (Figs.\ref{tunnel}b and \ref{summary}b), we find featureless flat tunneling conductance curves. When we find 2H-TaSe$_2$ (Figs.\ref{tunnel}c and \ref{summary}c) over different layers, clear superconducting tunneling features are observed with quasiparticle peaks located somewhat below $200$ $\mu$V. These features do not change as a function of the position over the surface, as shown by the zero bias conductance maps and its Fourier transform in Fig.\ref{summary}c .

\begin{figure}
\includegraphics[width=0.48\textwidth,clip]{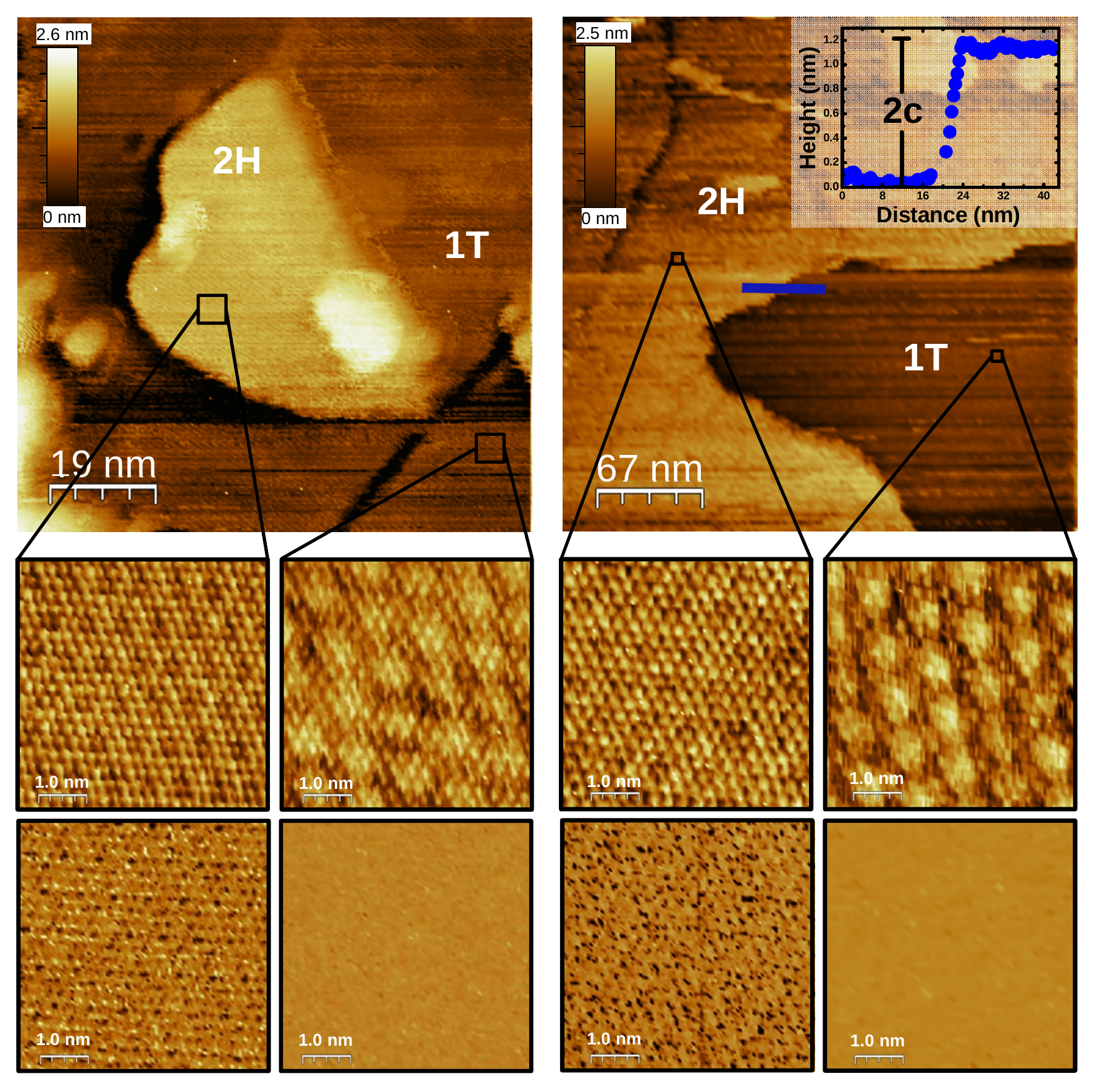}
\caption{Large topographic images in the top panels, and zoom-ups of different areas in the middle panels. The bottom panels are conductance maps taken at zero bias in the same areas as the small size topographic images shown in the middle panels, showing similar color scales as in Fig.\ref{summary}. We observe a 2H-TaSe$_2$ single layer on top of 1T-TaSe$_2$. The hexagonal layer shows the features in the conductance discussed in top panels of Figs.\ref{tunnel},\ref{summary}, and the trigonal layer is featureless. Images here are unfiltered and have been taken at 0.15 K. Height profile marked by a blue line in top right panel is shown in top right inset.}
\label{profile}
\end{figure}

In Fig.\ref{profile} we show representative examples of single layers of 2H-TaSe$_2$ imaged over larger areas (top panels), together with four zoom ups of topography (middle panels) and zero bias conductance maps (bottom panels) at the 2H single layers and at the 1T underlayer. Steps between single layer 2H-TaSe$_2$ correspond to the size of one unit cell of 2H-TaSe$_2$, around 1.2 nm. Typically, the lateral size of the single layer sheets are between 50 nm to 300 nm, and the boundary with the underlying 1T-TaSe$_2$ layer is sharp.

\begin{figure}
\includegraphics[width=0.46\textwidth,clip]{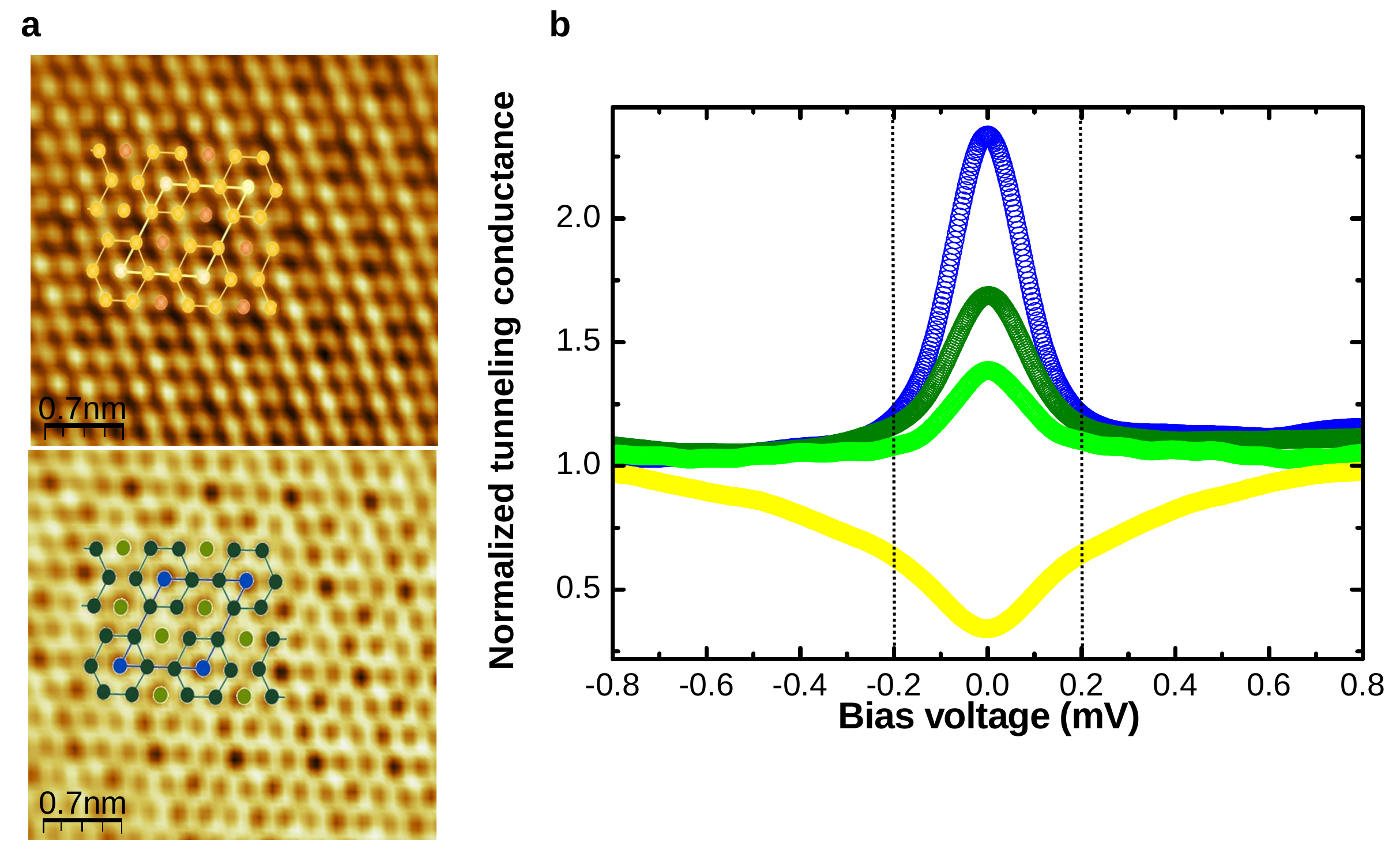}
\caption{Topography (top panel) and tunneling conductance map at zero bias (bottom panel) of a hexagonal layer on top of a trigonal layer. Images, taken at 0.15 K, are filtered for clarity. The blue, dark and light green curves of b are taken all three on top of Se atoms, corresponding to the color code shown in the bottom panel of a. The yellow curve is taken in between Se atoms. The curves on top of the Se atoms all show a zero bias conductance peak, whose height is modulated as a function of the position. Blue curves are located at the Se position which also shows highest contrast in the topography (white on top panel of a). Dark green and light green curves are located at the other Se positions with different charge modulations. Yellow curve is independent of the position in the CDW modulation.}
\label{peak}
\end{figure}

More detailed real space imaging of the 2H-TaSe$_2$ single layer is shown in Fig.\ref{peak}. The atomic and CDW modulations are observed simultaneously on the zero bias conductance map and the topography. In particular, the Fourier transform image of the zero bias conductance map shows the atomic Se lattice and the CDW. In real space, we observe that the highest value for the zero bias quasiparticle peak (blue curve in Fig.\ref{peak}b) coincides with the brightest Se atom due to the CDW (white points in Fig.\ref{peak}a). On the other hand, the V-shape dip (yellow curves in Figs.4a and 5a and in Fig.\ref{peak}b) between Se atoms is homogeneous. Thus, there is a very strong modulation of the zero bias peak at the Se atoms with the position related to the CDW order. The V-shape of the intersites shows no CDW modulation, and persists until bias voltages up to approximately 0.7 mV.

\begin{figure}
\includegraphics[width=0.48\textwidth,clip]{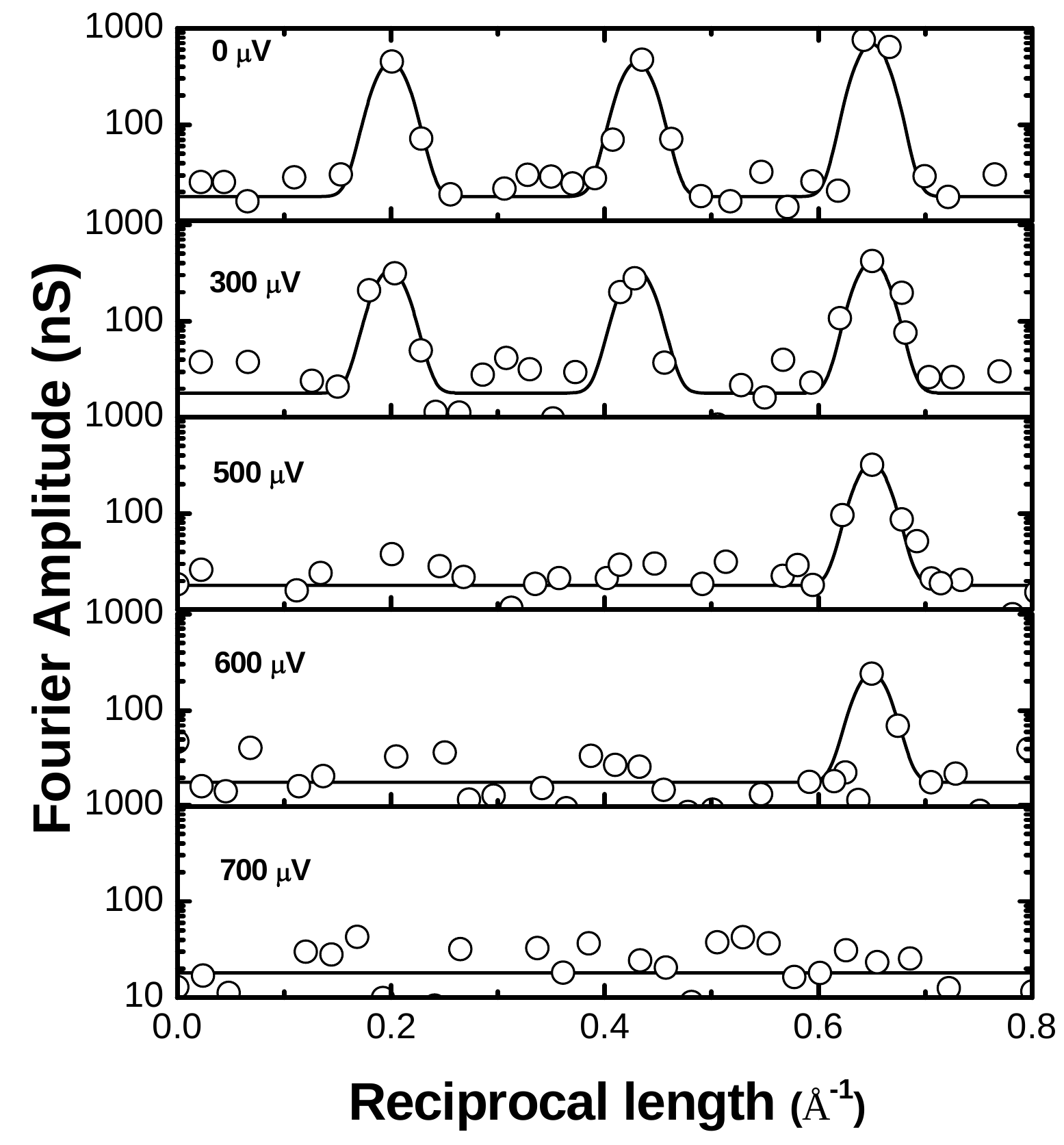}
\caption{The Fourier amplitude of the first three Bragg peaks in single layers of 2H-TaSe$_2$ on top of 1T-TaSe$_2$ along the direction of the six-fold modulation of highest contrast in the tunneling conductance maps. Note the use of log-scale to highlight the noise background. The central Bragg peak has been removed. The two peaks at the lower reciprocal lengths give the CDW pattern, and the third peak is due to the atomic Se lattice. When increasing the bias voltage, the sixfold modulation remains for the CDW peaks until it disappears above 300$\mu$V. Largest part of this modulation comes from the local variations in the zero bias peak amplitude. The modulation at distances of the atomic lattice (third Bragg peak) remains until 600$\mu$V. This is due to the modulation of the V-shape dip at intersites.}
\label{fourier}
\end{figure}

Accordingly, the conductance maps at bias voltages different from zero show a smooth evolution of the Moir\'e patterns presented in Fig.\ref{summary}a. The contrast related to the Se lattice and the CDW is shown in Fig.\ref{fourier}, where we plot the Fourier amplitude of the lattice and CDW peaks observed in the Fourier transform of the zero bias conductance map in Fig.\ref{summary}a for different bias voltages. The CDW peaks are maintained up to about 150$\mu$V, where they start to decrease. Above 300$\mu$V, all the atomic positions present similar conductance and only contrast between them and the intersites is observed in the conductance maps. This gives the six-fold modulation of the hexagonal lattice until roughly 0.7 mV, where all features in the conductance maps disappear. Thus, as shown in Fig.\ref{peak}, the zero bias conductance peak has a small energy scale of roughly 150$\mu$V and is linked to CDW, whereas the V-shape dip at Se atoms is broad and survives up to h
 igher bias voltages.

When we increase the temperature or the magnetic field, we observe that these features disappear from the tunneling conductance curves, which become flat above approximately 1 K (insets of Fig.\ref{temperature}). Above fields of a hundred millitesla, the peak and dip disappear into flat tunneling conductance curves. The weak coupling superconducting gap equation $\Delta=1.76k_BT_c$ gives, for T$_c$=1 K, $\Delta=$150$\mu$eV, which coincides with the width of the zero bias peak in the single hexagonal layer (Fig.\ref{peak}) and the size of the superconducting gap in multilayers (Fig.\ref{tunnel}c).

Similar temperature dependencies are observed for the superconducting gap measured in 2H-TaSe$_2$ multilayers (Fig.\ref{temperature}). Note that the superconducting gap observed in multilayers is strongly smeared, with a high number of states at zero bias, contrasting well developed gap structures found in the other 2H transition metal dichalcogenides\cite{Hess90,Guillamon08b,Guillamon08c,Guillamon11}.

Magnetic fields of 10 mT, applied perpendicular to the sample, lead to flat tunneling conductance curves both in single layers and in multilayers. Such fields are at the lower limit of our coil and magnet power supply system, and we did thus not follow the temperature dependence of the upper critical field nor study spatial gap dependencies under field.

\section{Discussion}

Our sample shows, in the bulk, clearly 2H-TaSe$_2$ features (Fig.\ref{transport}). However, on the surface we can find 1T-TaSe$_2$ CDW on some regions. Thus, the surface properties can significantly change with respect to the bulk. Tunneling conductance maps are featureless in 1T-TaSe$_2$ surfaces, which is in itself not very surprising, and shows that this polytype has no noticeable physics at energies of a mV or below. On the other hand, 2H-TaSe$_2$ surfaces are found with a critical temperature of 1 K. This is at odds with the superconducting transition found in the bulk (0.15 K). The critical temperature of these materials is easily enhanced through pressure or strain\cite{Bulaevskii76,Moncton77,Suderow05d,Coronado10}. The increase of T$_c$ highlights surface strains, or surface induced slight modifications in the electron-phonon coupling in some areas. Such modifications are probably more difficult to observe in other dichalchogenides such as NbSe$_2$ or NbS$_2$ where th
 e bulk T$_c$ is higher and closer to the maximum T$_c$ obtained in these materials under pressure, which lies around 9 K \cite{Suderow05d,Tissen13}. In our TaSe$_2$ samples, we did not observe any feature in the resistivity around 1 K. This means that the layers showing the superconducting gap of Fig.\ref{tunnel} (bottom panel) represent a very small volume fraction of the sample. The strongly broadened BCS features observed in Fig.\ref{tunnel} (bottom panel) also point that superconductivity is not that of a typical bulk and clean BCS s-wave superconductor.

At present, there are no clear-cut data of in-plane and out-of-plane coherence lengths of the bulk 2H-TaSe$_2$ superconductor with T$_c$=0.15 K. Values of $\xi_\parallel=500$nm (in-plane) and $\xi_\perp=200$nm (out of plane) have been obtained from the zero temperature extrapolation of the out-of-plane and in-plane critical fields (using $H_{c\perp}=\Phi_0/2\pi\mu_0\xi_{\parallel}^2$=1.4 mT and $H_{c,\parallel}=\Phi_0/2\pi\mu_0\xi_{\parallel}\xi_{\perp}$=4.1 mT) \cite{Yokota00}. However, resistivity measurements show an anisotropy of nearly three orders of magnitude (700), which is clearly at odds with the far smaller anisotropy found in the critical field measurements\cite{LeBlanc00}. Fermi surface measurements also point to strongly two-dimensional bands, so that the out of plane coherence length $\xi_{\perp}$ should be probably far below the value from critical field measurements \cite{Wilson75,Rossnagel11}. Regarding the in-plane coherence length $\xi_{\parallel}$, it is 
 strinking that the values given are much larger than those reported in 2H-NbSe$_2$, of 10 nm\cite{Suderow05d}. With so high values, the search for vortex lattice requires very large scanning ranges, above the size of the flat 2H-TaSe$_2$ areas observed in our experiment. Using the value we find here for the superconducting gap $\Delta$, we can make a simple estimation of the superconducting coherence length of multilayers of 2H-TaSe$_2$ with T$_c$ at 1 K and find  ($\xi_{\parallel}\approx\hbar$v$_F/2\Delta$ and v$_F=4.8\times10^{4}$m/s from Ref.\cite{Inosov09}) $\xi_{\parallel}\approx$ 500 nm, of similar order than the values discussed in the bulk\cite{Wilson75,Kumakura96,Yokota00}.

\begin{figure}
\includegraphics[width=0.5\textwidth,clip]{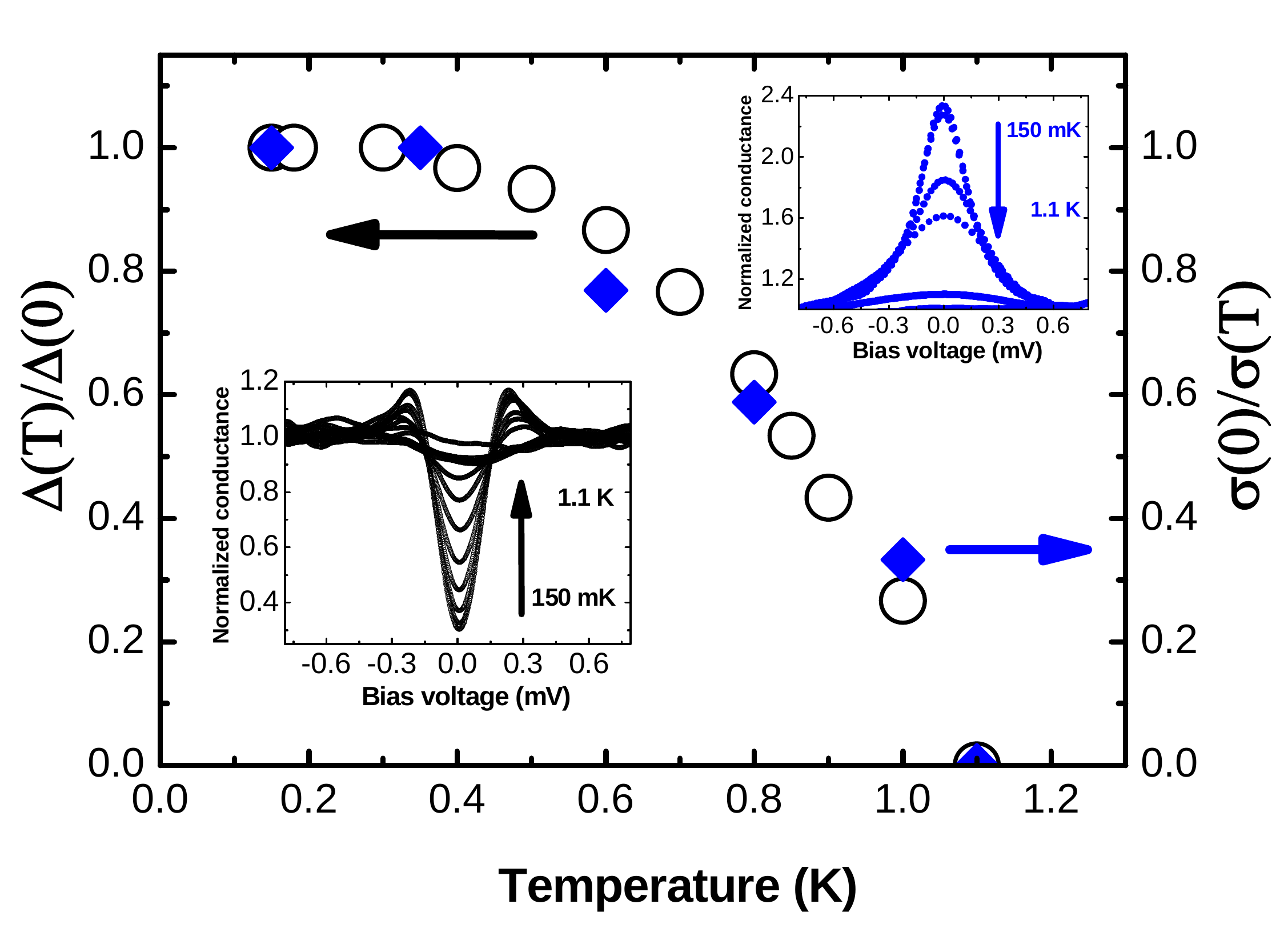}
\caption{The temperature dependence of the gap observed in multilayer 2H-TaSe$_2$  is shown as open black points at the left axis. The gap $\Delta(T)$ is normalized to its value at low temperatures. $\Delta(T)$ is obtained by fitting the curves shown in the lower left inset using a broadening parameter of 0.05 mV. The blue diamonds show the temperature dependence of the width of the Gaussian peak observed at Se sites, normalized to its value at low temperatures (0.1 mV) and inverted. Both quantities scale with each other, showing that the zero bias anomaly in single layers are related to the superconducting gap observed in multilayers. In particular, the broadening of the zero bias anomaly is not due to temperature increase, but to decreasing superconducting gap value.}
\label{temperature}
\end{figure}

The zero bias conductance peak found on single layers of 2H-TaSe$_2$ on top of 1T-TaSe$_2$ is probably our most intriguing result. We can compare (Fig.\ref{temperature}) the temperature evolution of the zero bias conductance peak in single layers with the evolution of the superconducting gap in multilayers. We can fit the tunneling conductance curves in multilayer 2H-TaSe$_2$ to s-wave BCS theory using a gap of $\Delta$=150$\mu$eV and a broadening lifetime parameter \cite{Dynes78} of $\Gamma=$55 $\mu$eV (black line shown in bottom panel of Fig.\ref{tunnel}). The temperature evolution of the gap parameter $\Delta(T)$ is shown by the open black points in Fig.\ref{temperature}. $\Delta(T)$ is below the nearly parabollic dependence expected within BCS theory, and it clearly disappears at 1 K. On the other hand, the zero bias conductance peak in single layer 2H-TaSe$_2$ follows well a Gaussian shape with a width of $\sigma$=0.1mV. When increasing temperature, the width of the peak
  increases above temperature induced broadening. The width of the peak in single layers inversely scales with the decrease of the gap in multilayers. This points that the origin of the zero bias conductance peak in single layers is related to superconductivity. It also shows that the destruction of superconducting correlations by temperature roughly follows $\Delta(T)$ in single layers, and the zero bias peak is correspondingly broadened by the temperature induced gap decrease.

\begin{figure}
\includegraphics[width=0.3\textwidth,clip]{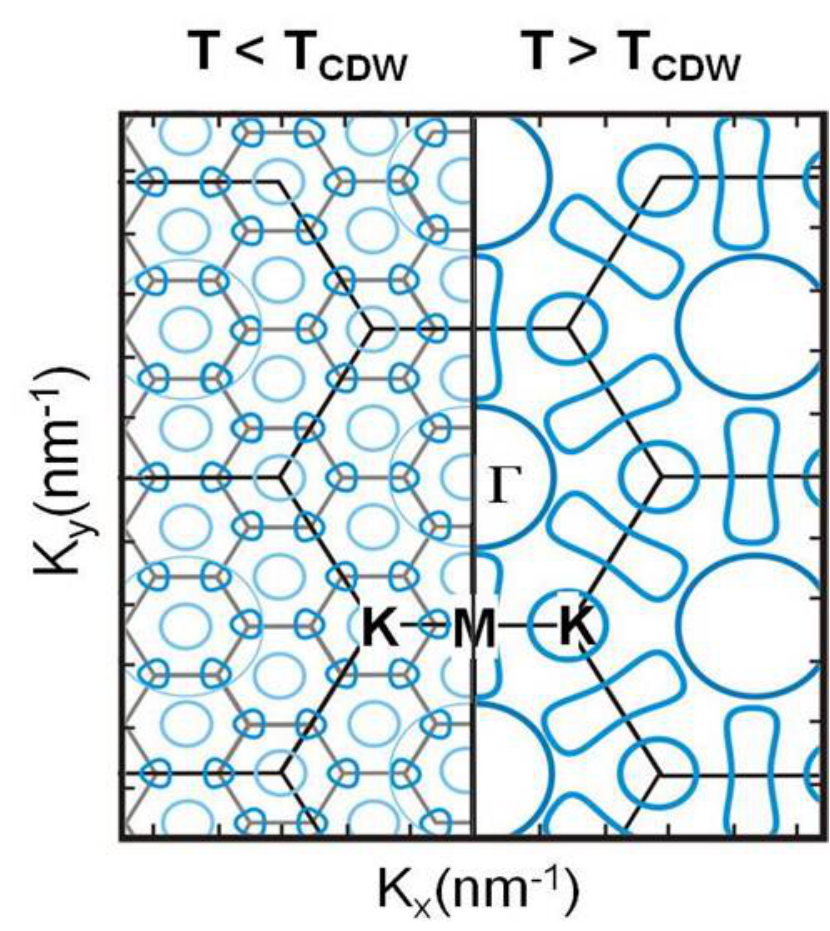}
\caption{Sketch of the Fermi surface above and below the charge density wave transition temperature. The sketch has been obtained by adapting figures of Ref.\protect\cite{Rossnagel11}.}
\label{fs}
\end{figure}

To discuss the atomic spatial dependence of the zero bias conductance peak, it is useful to mention the known features of the Fermi surface of 2H-TaSe$_2$ (Fig.\ref{fs}, see Ref.\cite{Rossnagel11}). At room temperature, it shows hole like sections centered at the $\Gamma$ and $K$ points, and electronlike dogbone-formed sheets around the $M$ point, coming from two different bands derived from Ta d electrons\cite{Borisenko08,Rossnagel11}). One for the $\Gamma$ and $K$ centered sheets, which has a saddle point in between, and the other one for the $M$ centered sheet. Angular resolved photoemission shows that small pseudogaps open on the $K$ pocket at high temperatures, precluding CDW order. At low temperatures, below the CDW transition, the Brillouin zone becomes three times smaller, and real bandgaps appear. The $\Gamma$ pocket remains intact, and the K pocket is destroyed, due to nesting features in the $\Gamma$ and $M$ sheets, with significant softening of $\Sigma_1$ phonons\
 cite{Moncton77}. The M dogbone sheet also suffers great changes. The low temperature Fermi surface consists of circular pockets at $\Gamma$ point of the new Brillouin zone and rounded triangles at the new $K$ points. The $M$ dogbone breaks up into parts. One is located around the $\Gamma$ and the other one around the $K$ point, both of which are new due to the $3\times3$ CDW state\cite{Borisenko08,Rossnagel11}. The strong spin-orbit coupling of the Ta 5d levels has a pronounced influence on the Fermi surface topology and could be at the origin of the gap on the dogbone sheets\cite{Rossnagel07}. There exists one part of the Fermi surface which remains untouched by the CDW order, and is strongly two dimensional, the big circle around $\Gamma$. In single layer 2H-TaSe$_2$, the CDW has the same structure than in the bulk, and thus Fermi surface, at least the part involved in charge order, is likely to have the same features too. 

Atomic size changes in the superconducting tunneling conductance have been often observed previously in systems with an anisotropic gap structure, such as 2H-NbSe$_2$ or the high T$_c$ cuprates\cite{Guillamon08PRB,Hoffman02a}. They result from the anisotropic interaction between the tip and the sample\cite{Guillamon08PRB}. The tunneling conductance probes the density of states of different parts of the band structure. The variations in the height of the zero bias peak with the CDW (Fig. \ref{peak}b) shows the involvement of charge order in shaping the zero voltage anomaly. Thus, the peak is related to the bands where CDW forms, i.e. the rounded triangles and the pockets stemming from the dogbone sheet.

Isolated single layer crystals of dichalcogenide materials have been obtained previously on different substrates through repeated exfoliation\cite{NatureNanotech12,Novoselov05,Novoselov05b,Castellanos10}. In spite of extensive searches, no clear experimental evidence of superconductivity has been found in them. On the other hand, in-situ grown single layer surfaces, mostly of Pb, are creating a rich playground, demonstrating that superconductivity can form in atomically thin crystals. In all cases, the substrate and the related interface plays a fundamental role, for instance the covalent bonding to a Si(111) substrate in sub-monolayers of Pb\cite{Zhang10}, or the interaction with the SrTiO$_3$ substrate in FeSe\cite{Liu12}. In any case, the superconducting critical temperature T$_c$ decreases when achieving ultimate thickness\cite{Qin09,Cren09,Zhang10}, and the superconducting tunneling conductance shows s-wave BCS like gap features\cite{Liu12}. Theory proposes significant e
 lectron-electron interactions in doped graphene layers\cite{Chubukov12,Profeta12}, or in single layer MoS$_2$\cite{Roldan13}, which give d-wave or superconductivity changing sign in different Fermi surface sheets.

Thus, both the observed strong critical temperature increase close to the surface, and the atomic size variation of the tunneling conductance between a zero bias peak and a V-shaped dip are first reported here. We can speculate of various possible origins for this behavior.

Regarding the zero bias conductance peak, zero energy resonances have been found previously in a number of systems, possibly receiving most attention in superconductors and in low dimensional structures with charging effects. Often, the zero bias peaks evidence the presence of a flat band.

In superconductors, resonances close to the Fermi level are found e.g. at the core of magnetic vortices because of multiple Andreev reflection\cite{Caroli64}. The first such state is located very close to zero energy (at $\Delta^2/E_F$, which is generally small) and has been seen in the STM experiments as a zero bias peak\cite{Hess90,Guillamon08b}. At zero magnetic field, in-gap bound states also arise at magnetic impurities in s-wave superconductors \cite{Yazdani97,Balatsky06,Franke12}. Their energy location depends on the relative amplitude of electron and hole impurity wavefunctions, which is governed by the exchange interaction between the localized magnetic moments of the impurity and the Cooper pairs, or by the scattering phase shift\cite{Balatsky06,Fischer07,Hoffman02a}. When scattering is resonant, the impurity bound state occurs exactly at the Fermi level, and the impurity spin is screened by an in-gap state oppositely polarized, in a similar way to Kondo screening i
 n a normal metal through singlet formation\cite{Balatsky06}. Magnetic properties have been found in gated MoS$_2$, but, in TaSe$_2$, there is until now no evidence for magnetic interactions. Thus, it seems difficult to discuss here bound states formed through magnetic scattering.

On the other hand, zero energy resonant bound states located exactly at the Fermi level arise in reduced symmetry superconductors, such as d-wave or more complex superconductors, when some kind of scattering leads to a sign change or a phase slip of the underlying wavefunction. A zero voltage conductance peak can appear thus close to impurities, at a surface or close to crystal boundaries \cite{Tanaka12,Tanaka96,Balatsky06,Pan00,Volovik11,Schnyder11,Stephanos11,Kopnin11,Chubukov12}. An intriguing possibility is that some sort of unconventional superconductivity appears within the 2H-TaSe$_2$ sheets, either in the form of d-wave order parameter, or of sign changing superconductivity between different sheets\cite{Mazin08,Roldan13}. Such a possibility would imply that the superconducting properties change from conventional s-wave in multilayers to unconventional reduced symmetry superconductivity when decreasing the thickness of the sample down to a single layer.

Resonant scattering gives sharp states with vanishing energy width. In the experiment, the observed peak is often broadened. Peak widths of several tenths of mV, being mostly smaller than the gap value, are found in many cases\cite{Balatsky06}.  For instance, the peak observed at the vortex core in 2H-NbSe$_2$ is about 1/3 of the gap value\cite{Hess90,Guillamon08b}. The zero bias conductance peak we observe here has a strong broadening, of the same order than the gap value. Broadening of zero bias conductance peaks in superconductors has been related to impurities, random disorder or complex gap variations over the Fermi surface \cite{Melnikov,Wimmer10,Balatsky06,Pikulin12}. We do not find evidence for impurities or random disorder. But gap size changes in different parts of the Fermi surface are likely to appear in the involved Fermi surface of 2H-TaSe$_2$ at low temperatures.

Other zero bias conductance peaks have been discussed in low dimensional structures, quantum dots and graphene. Bound states are formed by confinement or at interfaces, and, under appropriate conditions, these can lead to a zero energy state. For instance, edge states in graphite ribbons have been predicted to show flat bands near the edges\cite{Nakada96}. Electrostatic gating of graphene has been proposed to lead to confinement induced sharp peaks in the density of states\cite{Silvestrov07}. The combination of Coulomb blockade and quantum dots can exhibit zero bias anomalies\cite{Konig96}, and interface bound states have been predicted at the interface between graphene and superconductors\cite{Burset09,Kopnin11}. The role of interface and charging effects seems difficult to discuss in our single layers of 2H-TaSe$_2$ with the available data. The very recent discovery of superconductivity in gate-tuned MoS$_2$ devices shows that superconductivity can arise at the interface be
 tween a dichalcogenide, which is semiconducting in the bulk, and a substrate\cite{Ye12}.

The V-shaped conductance dip (yellow curves in Fig.\ref{tunnel}a) can be related to decreasing density of states due to superconducting correlations, in particular the decrease below 150$\mu$V. Such curves are clearly at odds with conventional BCS expressions, and highlight a rather peculiar density of states. They do not show any charge order related modulations, and could be thus reflecting behavior of the parts of the Fermi surface which are not affected by charge order. 

Let us remark that the cross-over between regions with widely different tunneling conductance occurs very sharply, just at atomic size at the step between both layers. This implies that there is significant electronic de-coupling between the top-most 2H-TaSe$_2$ and the substrate 1T-TaSe$_2$ layers. The 2H-TaSe$_2$ showing the zero bias peak acts as a separate single layer weakly coupled to its substrate. This, together with the multiband properties of the Fermi surface, could be the origin of peculiar electronic features. Separated layers may include, in particular, enough electron-electron repulsion in order to establish sign changing superconductivity\cite{Roldan13}.

In summary, we have found highly anomalous tunneling conductance features on the surface of 2H-TaSe$_2$. We observe a significant increase of the critical temperature close to the surface, and a zero bias peak in single layer crystals of 2H-TaSe$_2$ when they lie on top of a surface of 1T-TaSe$_2$. The zero bias peak is modulated by a charge density wave, and coexists with a V-shaped conductance dip. We do not fully understand the microscopic origin for the zero bias peak, but we have shown that it disappears at the same temperature where superconducting correlations disappear in multilayered 2H-TaSe$_2$. We briefly discuss different possibilities to explain the peak, including the presence of a resonant state at the Fermi level. Single layer properties of these materials have unexpected low energy features which are totally different from the bulk.

We acknowledge discussions with F. Guinea, V. Vinokur, T. Baturina, A.I. Buzdin and P. Monceau. We also acknlowledge advice and discussions about the dichalchogenides with J.L. Vicent. The Laboratorio de Bajas Temperaturas is associated to the ICMM of the CSIC. This work was supported by the Spanish MINECO (Consolider Ingenio Molecular Nanoscience CSD2007-00010 program, FIS2011-23488, MAT2011-25046 and ACI-2009-0905) and by the Comunidad de
Madrid through program Nanobiomagnet.

$^*$ Corresponding author, hermann.suderow@uam.es.


\end{document}